\documentclass[twocolumn,prl,showpacs,amsmath,amssymb,superscriptaddress]{revtex4}
\usepackage[T1]{fontenc}
\usepackage[latin1]{inputenc}
\usepackage{graphicx}

\begin{document}

\title{Geometric $\pi$ Josephson junction in $d$-wave superconducting thin films}

\author{A. Gumann}
\affiliation{Institut f\"ur Theoretische Physik, Universit\"at T\"ubingen, Auf der Morgenstelle 14, D-72076 T\"ubingen, Germany}
\author{C. Iniotakis}
\affiliation{Institute for Theoretical Physics, ETH Zurich, Schafmattstrasse 32, CH-8093 Zurich, Switzerland}
\author{N. Schopohl}
\affiliation{Institut f\"ur Theoretische Physik, Universit\"at T\"ubingen, Auf der Morgenstelle 14, D-72076 T\"ubingen, Germany}

\date{August 29, 2007}

\begin{abstract}
A novel way to realize a $\pi$ Josephson junction is proposed, based on a weak link in an unconventional
$d$-wave superconductor with appropriately chosen boundary geometry.
The critical current of such a junction is calculated from  a fully selfconsistent solution of 
microscopic Eilenberger theory of superconductivity. 
The results clearly show, that a transition to a $\pi$ Josephson junction occurs  for both low temperatures and small sizes of the
geometry.
\end{abstract}


\maketitle

Josephson junctions with an intrinsic phase shift of $\pi$ ($\pi$ Josephson junctions) open up promising possibilities in superconducting electronics. Including them in a closed superconducting loop allows to create a degenerate current ground state. 
If they are combined with standard Josephson junctions, complementary superconducting quantum interference devices (SQUIDs) with characteristics $I_c(\Phi)$ and $V_{dc}(\Phi)$ shifted by $\Phi_0/2$ with respect to the standard SQUID can be realized~\cite{Ter01} ($\Phi_0=h/2e$ is the flux quantum). These devices permit various improvements in rapid single flux quantum logic (RSFQ,~\cite{Lik01}). Using $\pi$ Josephson junctions, complementary logic devices can be realized without the need for additional current bias lines, with improved device symmetry and enhanced operation margins~\cite{Ort01}. At the same time, the size of the logic cells can be significantly reduced~\cite{Ust01}.
\par
The first proposal for a $\pi$ Josephson junction was based on a tunnel junction with magnetic impurities~\cite{Bul01}. Experimental realizations in the form of superconductor-ferromagnet-superconductor~\cite{Rya01,Bau01} or superconductor-isolator-ferromagnet-superconductor~\cite{Kon01,Blu01,Wei02} mulitlayered systems have been presented.
A second realization of $\pi$ Josephson junctions makes use of the $d_{x^2-y^2}$ ($d$-wave) pairing symmetry of the cuprates via grain boundaries intersecting domains with different orientation of the crystal lattice~\cite{Tsu01,Bar01,Schu01,Il02}. Another  method exploits the pairing symmetry of the cuprates by combining high-T$_c$ and low-T$_c$ materials~\cite{Wol01,Bra01,Smi01}. Furthermore, $\pi$ Josephson junctions have been realized in superconductor-normal conductor-superconductor structures with a nonequilibrium energy distribution of the current-carrying states in the normal region~\cite{Bas01}.
\par
\begin{figure}
\begin{center}
\includegraphics[width=0.99\columnwidth, keepaspectratio]{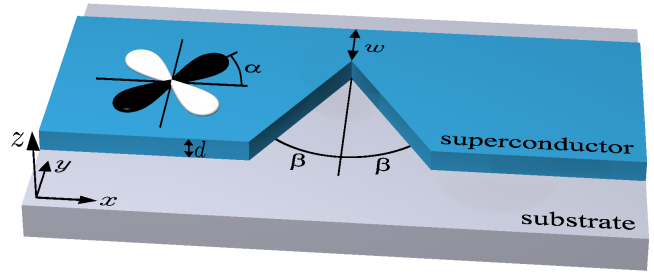}
\end{center}
\caption{\label{cap:geometry}(Color online) Geometry of the $\pi$ Josephson junction based on a thin film of a  $d$-wave superconductor.}
\end{figure}

In this letter, we propose a novel realization of a $\pi$ Josephson junction, which is solely based on the boundary geometry of a $c$-axis oriented $d$-wave superconductor thin film. We consider an epitactic film of the superconducting material exhibiting a weak link of width $w$
as displayed in Fig.~\ref{cap:geometry}. The geometry of this weak link consists of a straight line on one side and a
wedge-shaped  incision of angle $2\beta$ on the other. 
We point out that the weak link defined by this geometry is a $\pi$ Josephson junction  if (1)~the crystal orientation is appropriately chosen and (2)~the residual width $w$ is sufficiently small.
\par
The intrinsic phase shift of $\pi$ following from the proposed geometry  is a direct consequence of the unconventional $d$-wave pairing symmetry. If the orientation angle between $d$-wave and geometry is chosen to be $\alpha=\pi/4$ as indicated in Fig.~\ref{cap:geometry}, quasiparticles travelling through the constriction, which get reflected at the straight boundary line (opposite to the wedge), simultaneously suffer a sign change of the pairing potential. Thus, they generate zero energy Andreev bound states~\cite{Hu01} in the junction associated with a phase shift of $\pi$. In contrast, quasiparticles passing through the constriction without such a reflection do not contribute to the zero energy Andreev bound states. If the residual width $w$ is sufficiently small, however, the contribution of the reflected quasiparticles dominates the total current across the junction resulting in a $\pi$ Josephson junction behaviour.

In the following, we calculate the critical current $I_c$ of the junction according to  Fig.~\ref{cap:geometry},
as a function of width $w$ and temperature $T$.
For this purpose, we solve the Eilenberger equations of superconductivity~\cite{Eilenberger,Larkin} to take into account the effect of Andreev bound states quantitatively.
We assume a cylindrical Fermi surface of the superconductor, which is  aligned parallel to the $z$ axis and parametrized
by the polar angle $\theta$. Accordingly, the Fermi velocity is given by 
$\mathbf{v}_F=v_F(\hat{\mathbf{x}}\cos{\theta}+\hat{\mathbf{y}}\sin{\theta})$. 
The pairing potential in the superconductor may be factorized
as $\Delta(\mathbf{r},\hat{\mathbf{k}})= \psi(\mathbf{r}) \chi(\hat{\mathbf{k}})$
with  $\chi(\hat{\mathbf{k}})=\cos(2\theta-2\alpha)$ representing the $d$-wave symmetry. Then,
the selfconsistency equation for the pairing potential according to Eilenberger theory is given by
\begin{equation}
\label{eq:gapeq}
\psi(\mathbf{r})=
2\pi N(0) V k_B T\sum_{\varepsilon_n>0}^{\omega_c}\big<\chi(\hat{\mathbf{k}})f(\mathbf{r},\hat{\mathbf{k}},i\varepsilon_n)\big>_{\theta},
\end{equation}
and the current density can be computed from
\begin{equation}
\label{eq:curreq}
\mathbf{j}(\mathbf{r})=
4\pi e N(0) k_B T\sum_{\varepsilon_n>0}^{\omega_c}\big<\mathbf{v}_F\cdot g(\mathbf{r},\hat{\mathbf{k}},i\varepsilon_n)\big>_{\theta}.
\end{equation}
In these equations $N(0)$ is the normal density of states at the Fermi surface, $V$ is the coupling constant, $\varepsilon_n=(2n+1)\pi k_B T$ are Matsubara frequencies, and $\langle ... \rangle_\theta$ denotes Fermi surface averaging.
The propagators $f$ and $g$ in the integrands of Eqs.~(\ref{eq:gapeq}) and~(\ref{eq:curreq}), respectively, 
can easily be calculated using the Riccati parametrization~\cite{Schop02,Schop01}.
\par
We numerically solve the selfconsistency Eq.~(\ref{eq:gapeq}) in the $xy$ plane for a two-dimensional section of the superconducting thin film enclosing the constriction shown in  Fig.~\ref{cap:geometry}. This section has an area of more than $12.5\xi_0\times12.5\xi_0$, and a grid width of about $0.15\xi_0$ is used ($\xi_0=\hbar v_F/\pi\Delta(T\!=\!0)$ is the coherence length). All the boundaries of the geometry are assumed to be impenetrable, leading to specular reflection conditions which are incorporated appropriately~\cite{She01}. The left and right sides are considered to be open ends, exhibiting a fixed phase difference $\Delta\phi$ between them as a constraint. Details of the iterative numerical procedure which we use in order to find the selfconsistent solution can be found in Ref.~\cite{Gum01}. In order to find the critical current, these solutions have to be calculated for a large number of phase differences. In a second step, the current Eq.~(\ref{eq:curreq}) is used to access the whole  current-phase relation of the junction. Finally, the critical current $I_c$ is extracted as the absolute maximum of the current-phase relation.

\begin{figure}
\begin{center}
\includegraphics[width=0.99\columnwidth, keepaspectratio]{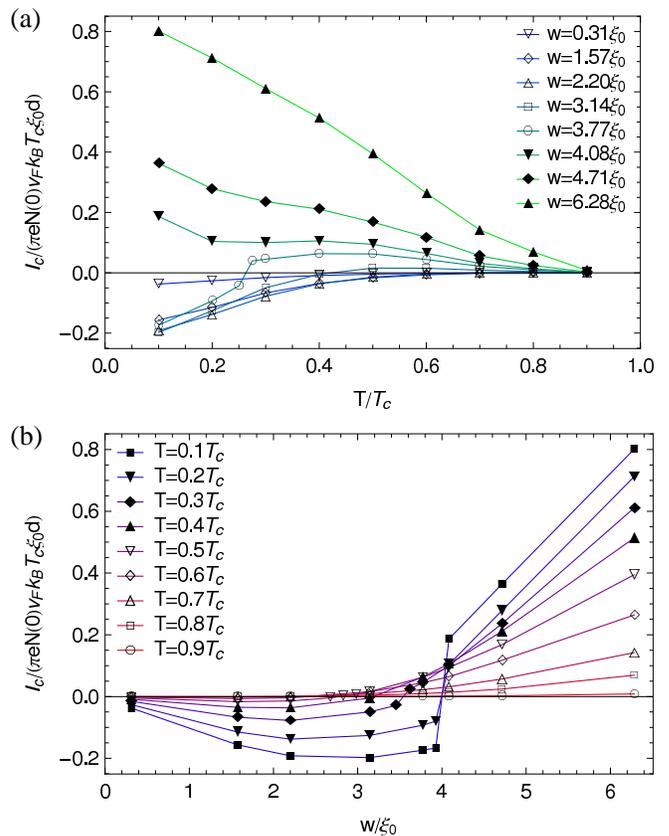}
\end{center}
\caption{\label{cap:Ic0bT}(Color online) Critical current $I_c(w,T)$ for $\alpha=\pi/4$ and $\beta=0$. The two plots (a) and (b) correspond to two experimental situations, which can be thought of to verify the $0$-$\pi$-transition, i.e. by variation of  $T$ or $w$, respectively. The lines are guides for the eye.}
\end{figure}

\par
In Fig.~\ref{cap:Ic0bT}, we present our results for  the critical current $I_c(w,T)$ of the weak link for $\alpha=\pi/4$ and $\beta=0$. Here, $T_c$ is the transition temperature and $d$ is the thin film thickness. In~Fig.~\ref{cap:Ic0bT}(a), the data is plotted for fixed width $w$ over temperature $T$, whereas, in~Fig.~\ref{cap:Ic0bT}(b), the same data is plotted for fixed temperature $T$ over width $w$.
The results in Fig.~\ref{cap:Ic0bT}(a) show that the $\pi$ state (indicated by a negative critical current) is predominantly entered at low temperatures. For very small values of $w$ however, the $\pi$ state survives up to the highest temperatures that we considered ($0.9\,T_c$), featuring small absolute values of the critical current. From the results in Fig.~\ref{cap:Ic0bT}(b), we find that at $T=0.1\,T_c$, the $0$-$\pi$-transition occurs at about $b \simeq 4\xi_0$. With increased temperature, the critical width of the $0$-$\pi$-transition is shifted to smaller values of $w$.
\par
For the geometry $\alpha=\pi/4$ and $\beta=\pi/4$, we find similar results with a $0$-$\pi$-transition at slightly smaller values of $w$ (not shown here). This can easily be understood since only quasiparticles  from a reduced angular interval contribute to the $\pi$ state. Nevertheless, the occurence of the $\pi$ state only weakly depends on the angle of the wedge $\beta$.
\par
Based on our results for the critical current $I_c(w,T)$, we expect the experimental realization of the proposed Josephson device to be challenging, but feasible. The size of the coherence length of the superconducting material directly corresponds to the necessary size of the structures. As stated above, the occurence of the $\pi$ state of the proposed Josephson junction hardly depends on the angle of the wedge $\beta$. Furthermore, previous studies indicate that the occurence of surface Andreev bound states in $d$-wave superconductors is not suppressed by microscopic surface roughness~\cite{Ini01}. Accordingly, also the $\pi$ state of the Josephson junction proposed 
here should exhibit some robustness regarding surface roughness.
\par

\begin{figure}
\begin{center}
\includegraphics[width=0.99\columnwidth, keepaspectratio]{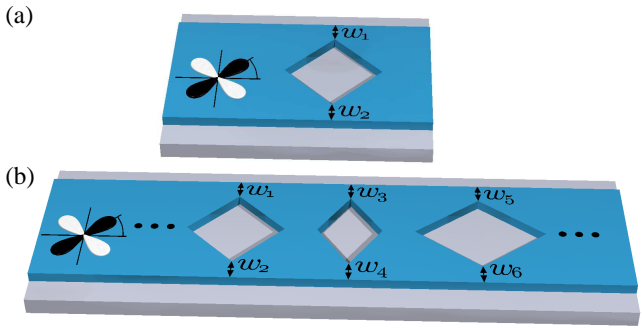}
\end{center}
\caption{\label{cap:squidsqif}(Color online) In (a), a combination of two of the proposed Josephson devices to a SQUID is shown. Since the widths of the two junctions $w_1$ and $w_2$ can be chosen independently, a $0$-$0$-, $0$-$\pi$- or $\pi$-$\pi$-SQUID is viable. In (b), a larger number of the proposed Josephson devices is combined to a SQIF (6 junctions shown, more indicated by the black dots).}
\end{figure}

The proposed realization of  $\pi$ Josephson junctions based on the boundary geometry of $d$-wave superconducting thin films opens up interesting possibilities for application. It allows for the fabrication of $0$ and $\pi$ Josephson junctions with similar characteristics in the same process. The material can be epitactic thin films of any superconductor with $d$-wave pairing symmetry. A cuprate high temperature superconductor can be employed as well as a $d$-wave heavy-fermion superconductor like CeCoIn$_5$~\cite{Mat01}. Because of the simple planar geometry, the combination of two of the proposed Josephson devices to $0$-$0$-, $0$-$\pi$- or $\pi$-$\pi$-SQUIDs is straightforward. An example for a possible geometry of such a device is shown in Fig.~\ref{cap:squidsqif}(a). Furthermore, the application of a large number of the proposed Josephson devices for superconducting quantum interference filters (SQIFs,~\cite{Schu02}) containing both $0$-$0$- and $0$-$\pi$-SQUID loops offers new possibilities in the synthesis of the voltage response of such a device. A possible geometry for a serial SQIF is shown in Fig.~\ref{cap:squidsqif}(b).
\par
Depending on temperature $T$ and the width of the junction $w$, the critical current of the proposed Josephson device can be comparatively small. This may pose a problem in some applications, but can be advantageous in others. In the case of a single SQUID, a small critical current implies a small value of the $\beta_L$ parameter. Thus, SQUIDs with larger dimensions can be designed, leading to an increased sensitivity. In the case of a serial SQIF, whose total voltage response is the sum of the voltage outputs of a large number of single SQUIDs, lower noise figures are expected.
\par
In conclusion, we proposed a new way to realize a $\pi$ Josephson junction, just consisting of
a single layer of a $d$-wave superconductor thin film with appropriately chosen boundary geometry. 
A fabrication of this geometric $\pi$ Josephson device is in the reach of modern fabrication technology. This method may allow to create more evolved  
superconducting circuits containing both normal and $\pi$ Josephson junctions
closely packed on a single substrate.

\end{document}